\documentclass[twocolumn,twoside,slac_two]{revtex4}
\usepackage{graphicx}
\usepackage{fancyhdr}
\begin{document}

\title{Initial Results of New Tomographic Imaging of the Gamma-Ray Sky with BATSE}

\author{G.~L.~Case, M.~L.~Cherry}
\affiliation{Dept.~of Physics and Astronomy, Louisiana State University, Baton Rouge, LA 70803, USA}
\author{J.~C.~Ling, M.~Lo, T.~Shimizu}
\affiliation{Jet Propulsion Laboratory, California Inst. of Technology, Pasadena, CA 91109, USA}

\author{Wm.~A.~Wheaton}
\affiliation{Infrared Processing and Analysis Center, California Inst. of Technology, Pasadena, CA 91125, USA}

\begin{abstract}
We describe an improved method of mapping the gamma-ray sky by applying the Linear Radon Transform to data from BATSE on NASA's {\it CGRO}. Based on a method similar to that used in medical imaging, we use the relatively sharp ($\sim 0.25^{\circ}$) limb of the Earth to collimate BATSE's eight Large Area Detectors (LADs).  Coupling this to the $\sim 51$-day precession cycle of the {\it CGRO} orbit, we can complete a full survey of the sky, localizing point sources to $<1^{\circ}$ accuracy.  This technique also uses a physical model for removing many sources of gamma-ray background, which allows us to image strong gamma-ray sources such as the Crab up to $\sim 2$ MeV with only a single precession cycle. We present the concept of the Radon Transform technique as applied to the BATSE data for imaging the gamma-ray sky and show sample images in three broad energy bands (23-98 keV, 98-230 keV, and 230-595 keV) centered on the positions of selected sources from the catalog of 130 known sources used in our Enhanced BATSE Occultation Package (EBOP) analysis system.  Any new sources discovered during the sky survey will be added to the input catalog for EBOP allowing daily light curves and spectra to be generated.  We also discuss the adaptation of tomographic imaging to the Fermi GBM occultation project.
\end{abstract}

\maketitle

\thispagestyle{fancy}

\section{Introduction}
The soft gamma-ray sky in the $200-2000$ keV range is one of the more poorly studied regions of the electromagnetic spectrum.  Many of the most intrinsically interesting objects in the universe emit much of their primary energy in this regime.  The last complete all-sky survey above $\sim100$ keV was made by {\it HEAO} in the late 1970s and early 1980s. While missions such as {\it INTEGRAL} have considerably better angular resolution and sensitivity than {\it HEAO} A4, they have a small field of view and reduced sensitivity above 200 keV.  Because of the penetrating power of gamma rays, massive collimators would be needed in order to achieve reasonable sensitivity in the $\sim 200-2000$ keV region.

\begin{figure}[b]
\includegraphics[width=3.25 in]{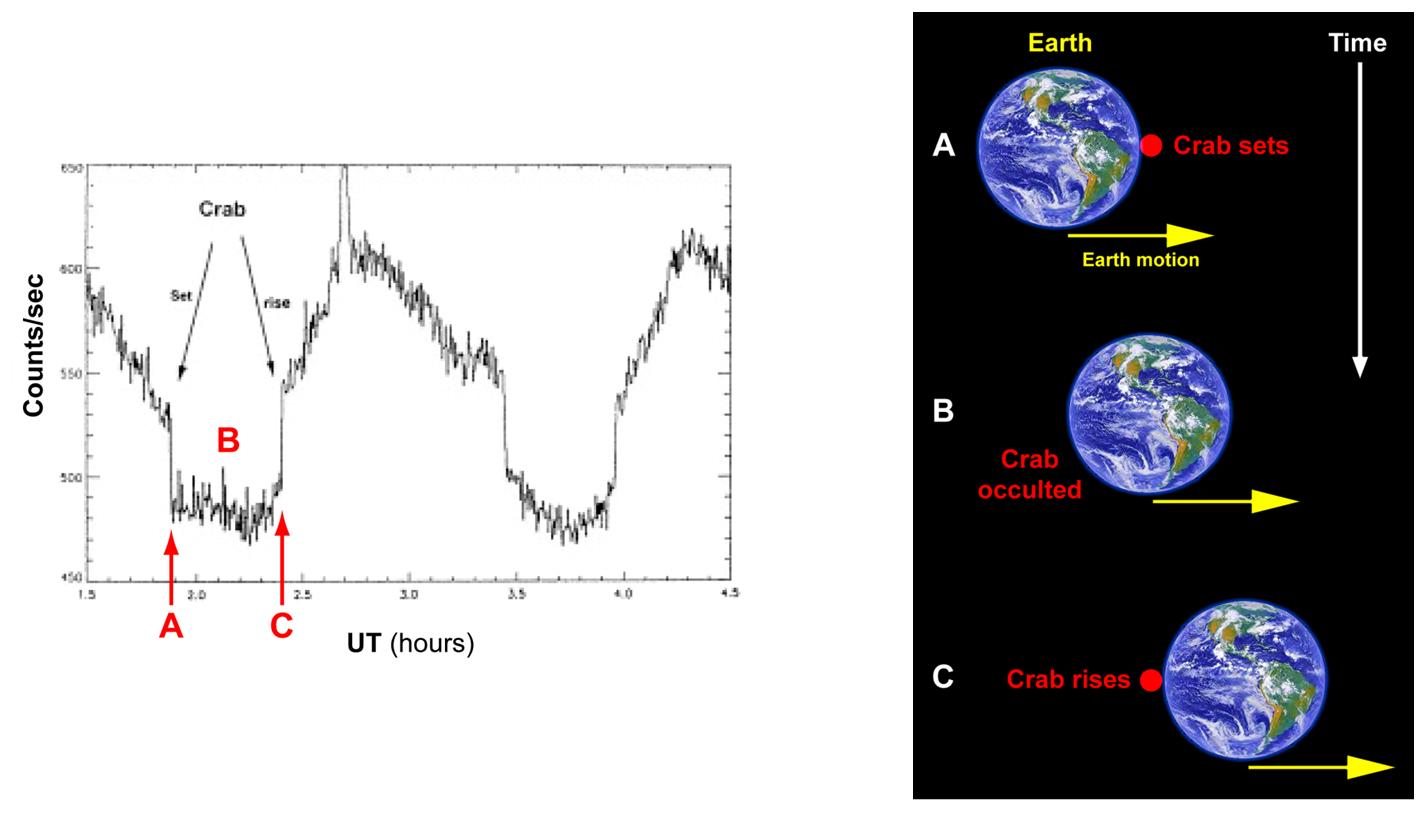}
\caption{Left: Time history from a BATSE LAD.  `A' marks the time at which the Crab is occulted, `B' marks the time during which the Crab is blocked by the Earth, and `C' marks the time the Crab reappears. Right:  Schematic of the Earth occultation process.}
\label{fig:occ}
\end{figure}

The Burst And Transient Source Experiment (BATSE) instrument was one of the four instruments on the {\it Compton Gamma Ray Observatory} ({\it CGRO}).  It consisted of 8 NaI large area detectors (LADs), 50 cm in diameter and 1.27 cm thick, each paired with a Spectroscopy Detector, 12.7 cm in diameter and 7.62 cm thick, located on the corners of the spacecraft.  Although the LADs have no direct imaging capability, it is possible to monitor known gamma-ray sources using the Earth occultation technique \cite{ling00,harmon02}.  The Earth occultation technique utilizes the relatively sharp ($\sim 0.25^{\circ}$) limb of the Earth to collimate photons from individual sources. Figure~\ref{fig:occ} demonstrates how the occultation technique works.  The count rate measured by a LAD is comprised of contributions from all of the cosmic sources not occulted by the Earth as well as a background component.  When a source is occulted by the Earth, the count rate measured in the LAD drops, producing a downward step-like feature in the count rate.  When the source reappears, an upward step is produced.  

The Enhanced BATSE Occultation Package (EBOP) was developed at JPL \cite{ling00} to fit the continuous count rate (CONT) data for an entire day to a model including all of the occultation steps that occur for the sources in the input catalog along with a semi-physical background model. The output consists of the average daily count rates in each LAD for each source in each of 16 energy bands.  However, EBOP requires {\it a priori} knowledge of the positions of the gamma-ray sources, i.e. a predetermined input catalog.  The discovery of new sources must be accomplished using other methods, such as imaging.

\begin{figure*}[t]
\includegraphics[width=6.0 in]{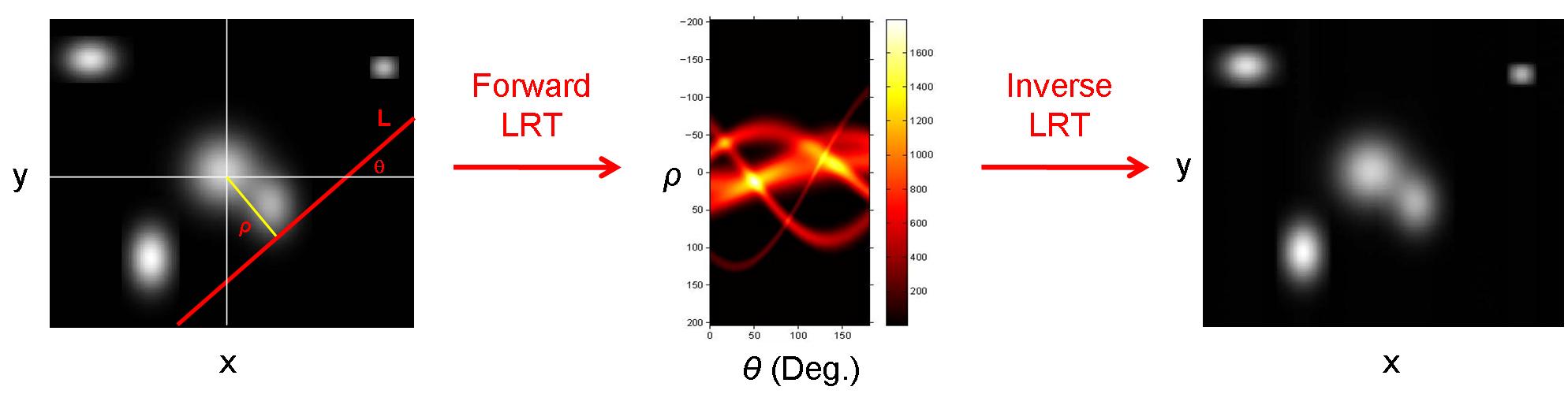}
\caption{Results of the forward and inverse Linear Radon Transforms.}
\label{fig:lrt}
\end{figure*}

Using BATSE, images of the sky in the $25-2000$ keV energy range can be obtained utilizing tomographic imaging techniques such as the Linear Radon Transform, a technique used in medical imaging applications.  Applying this technique to the BATSE CONT data using the Earth limbs gives an angular resolution roughly equal to the $\sim15'$ size of the atmosphere, as viewed from {\it CGRO}, that the source must traverse as it occults. With nine years of data available, images of the entire sky can be generated with high sensitivity, and the BATSE dataset can be used to search for previously unknown gamma-ray sources.

\section{The Linear Radon Transform}
The Linear Radon Transform (LRT) is the integral transform of an image $I$ across all lines $L$ in the $(x,y)$ plane into the $(\rho,\theta)$ plane, where $\theta$ is the angle that the line $L$ makes with respect to the x-axis and $\rho$ is the perpendicular distance from the line to the origin, 
\begin{equation} \label{eq:lrt}
R(I) = \int I(x,y) ~dL = F(\theta,\rho), 
\end{equation}
where $L = \left\{ y = x \tan \theta + \rho/\cos \theta \right\}$.

In Fig.~\ref{fig:lrt}, the LRT and its inverse are demonstrated for a sample image.  The middle image shows how the features in the $(x,y)$ plane are transformed into curves in the $(\rho,\theta)$ plane.  The Inverse LRT transforms the integrated curves back into the $(x,y)$ plane.

This technique was first applied to Earth occultation using BATSE data below 100 keV by Zhang et al.~\cite{zhang93}. We implement it as follows.  The 2.048-second CONT data is rebinned to 16.384 seconds and the background model, as obtained with EBOP, is subtracted.  The difference in the counts measured by a detector in two adjacent time bins, $t_0$ and $t_1$ with $t_1 > t_0$ for rising steps and $t_1 < t_0$ for setting steps, gives the integrated number of counts from the arc defined by the position of the Earth's limb at time $t_0$.  Thus an Earth occultation measurement is a LRT of the sky across a single``line''.  The LRT assumes a straight line, so the image tile must be restricted to small enough regions of the sky that the Earth's limb can be approximated as a straight line.  As the plane of the CGRO orbit precesses (with a period of $\sim51$ days given its $28.5^{\circ}$ inclination), the Earth's limb will be oriented at different angles with respect to the image tile.  An image is produced by performing the Inverse LRT on all of the limbs in one precession period.

\section{The Earth Limbs and the Point Spread Function} \label{sec:psf}
\begin{figure*}
\label{fig:psf}
\includegraphics[width=6.5 in]{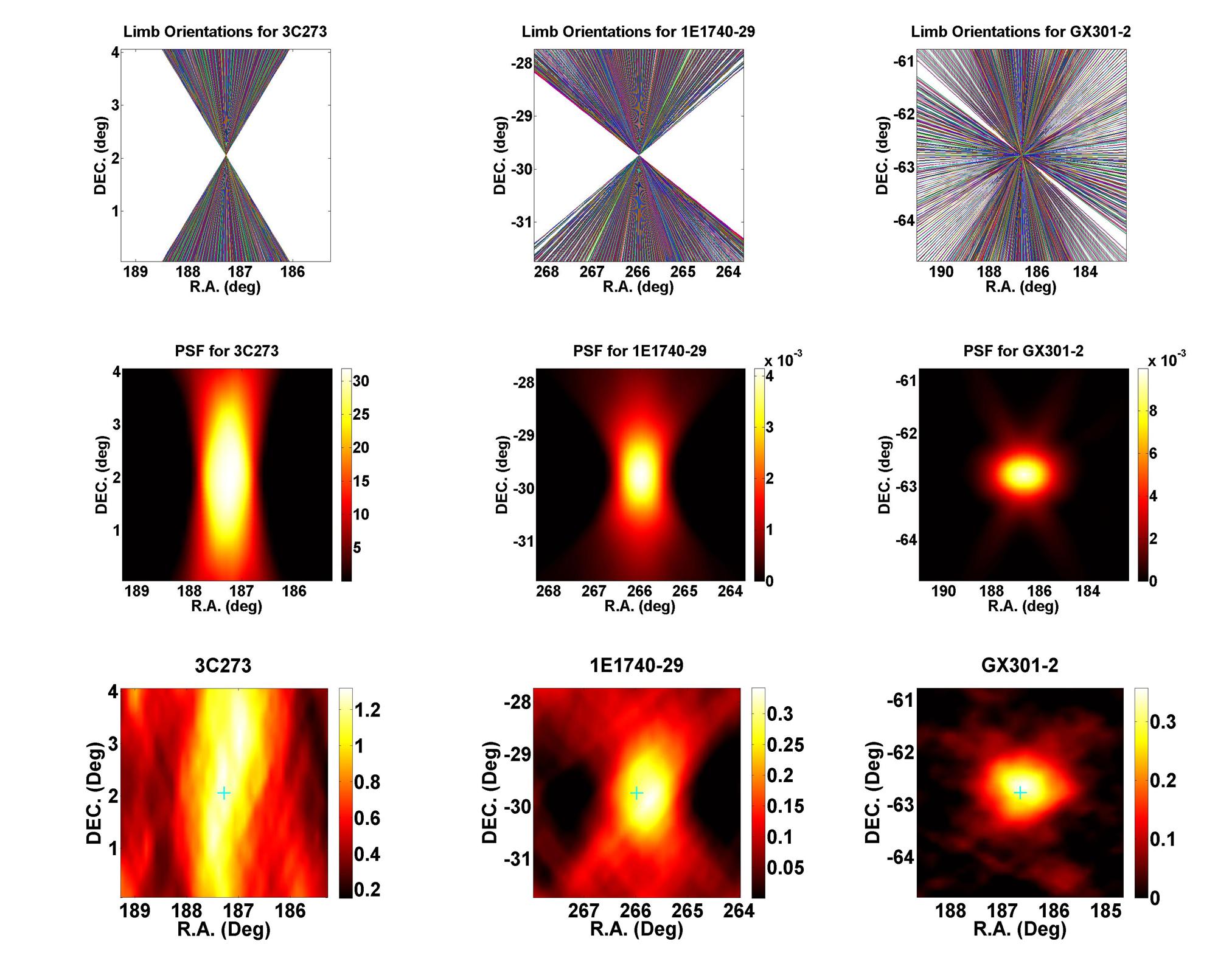}
\caption{Top: Orientation of limbs during one precession cycle for three sources at different declinations.  The colors of the lines have no meaning and are used to distinguish individual lines.  Middle: Simulated PSFs for the same three sources.  Bottom:  Actual images in the 23-98 keV band of the three sources averaged over 20 precession cycles.}
\end{figure*}

The point spread function (PSF), and hence the expected shape of the image, depends on the declination of the source.  Figure~\ref{fig:psf} shows the limb orientations (drawn as multi-colored lines) that are sampled during a complete orbit plane precession cycle for sources at three different declinations, along with the simulated PSFs and the actual images.  For a source at low declination (e.g.~3C273), fewer limb orientations are sampled (that is, the lines $L$ have a smaller range of angles, $\theta$) resulting in a PSF that is elongated in the north-south direction.  For sources at modest declination (e.g.~1E 1740-29), about half of the possible limb orientations are sampled, resulting in a PSF that is slightly elongated in the north-south direction with an `x' pattern in the wings.  For sources with declination $ \left| \delta \right| > 45^{\circ}$ (e.g.~GX 301-2), nearly all of the limb orientations are sampled and the PSF is roughly circular or even elongated in the east-west direction.  The PSFs are calculated by simulating the response to a unit intensity source for a particular time period, taking into account the expected limb orientations and the gaps that occur in the real data.  Gaps can appear in the data due to events such as passes through the South Atlantic Anomaly (when the detector high voltage is turned off), gamma-ray bursts, transmission dropouts, etc. As can be seen in Fig.~\ref{fig:psf}, the shapes of the images agree well with the simulated PSFs.

\section{Images}
\begin{figure*}
\includegraphics[width=6.5 in]{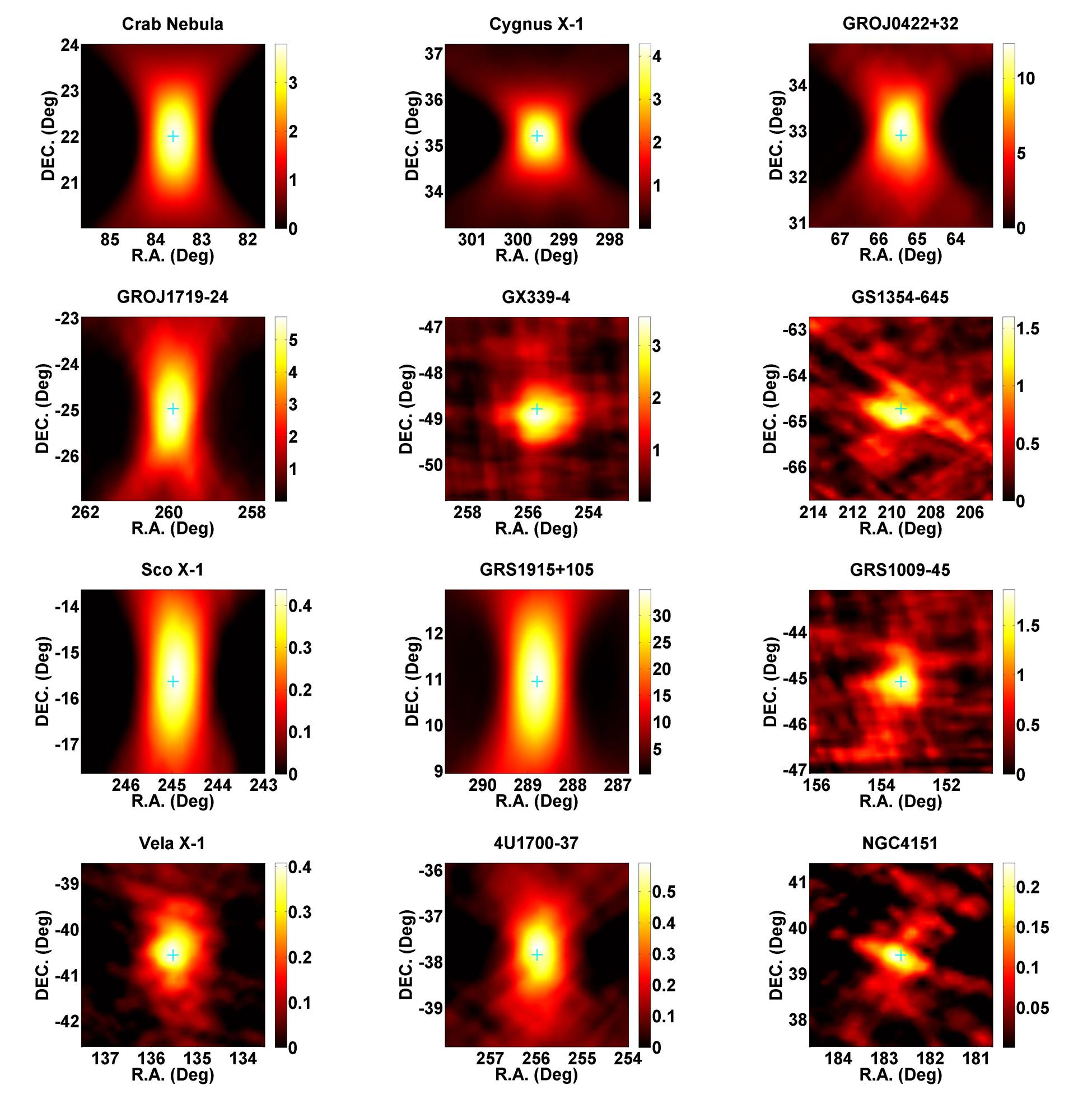}
\caption{Images in the 23-98 keV band (band 1). The intensity units are arbitrary and the `+' sign marks the known positions of the sources.}
\label{fig:eb1}
\end{figure*}

\begin{figure*}
\includegraphics[width=6.5 in]{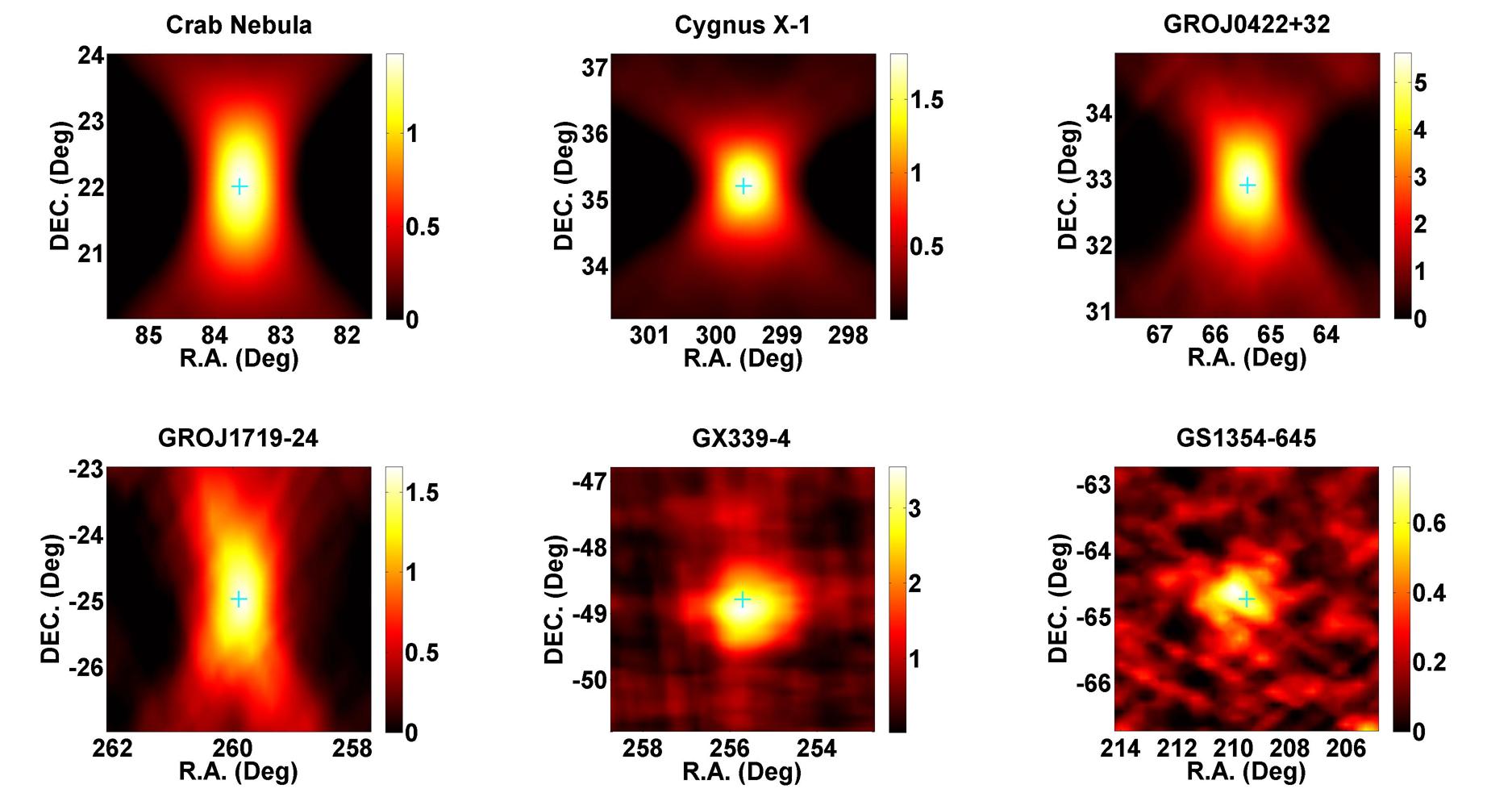}
\caption{Images in the 98-230 keV band (band 2).  The intensity units are arbitrary and the `+' sign marks the known positions of the sources.}
\label{fig:eb2}
\end{figure*}

\begin{figure*}
\includegraphics[width=6.5 in]{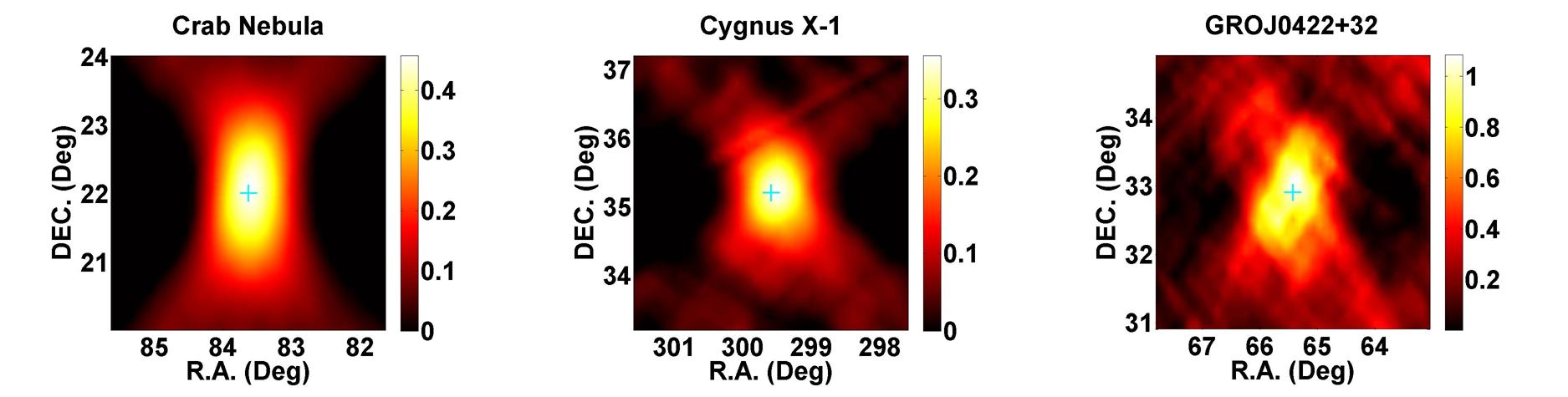}
\caption{Images in the 230-595 keV band (band 3).  The intensity units are arbitrary and the `+' sign marks the known positions of the sources.}
\label{fig:eb3}
\end{figure*}

Figure~\ref{fig:eb1} shows a selection of images generated with the Inverse LRT in the 23-98 keV (band 1) energy band, while Fig.~\ref{fig:eb2} and Fig.~\ref{fig:eb3} show images from the same time period for the 98-230 keV (band 2) and 230-595 keV (band 3) energy bands, respectively.  The images for persistent sources are summed over 20 precession periods, from TJD 8393-9397, while transient sources are summed over one or more complete precession cycles spanning the time of the outburst. Complete precession cycles are used in order to make sure that all available limb orientations are sampled, which creates a more uniform image.  All detectors that view the source at an angle $\phi < 60^{\circ}$ from the detector normal are combined, weighted by the detector efficiency. 

The Crab pulsar/nebula is the brightest source seen by BATSE in all energy bands.  It is a persistent source and is clearly seen in all three energy bands, even over a single precession cycle.  In fact, the Crab is also seen in band 4 (595-1800 keV) \cite{ling10}.  Cyg X-1, a high-mass x-ray binary harboring a black hole, is a persistent but variable source that is also seen clearly in all three energy bands.  GRO J0422+32 is a transient source powered by a black hole that was discovered by BATSE, and at its peak was several times brighter than the Crab in band 1.  During its outburst, a hard spectral component was observed by Ling \& Wheaton \cite{ling03}, which is evident in the band 3 image.  

GRO J1719-24 is another transient black hole source discovered by BATSE.  During its outburst, it also was observed to have a hard spectral component \cite{ling05} and is about the same intensity as the Crab in band 2.  GX 339-4 is also a transient black hole source that underwent four outbursts from 1991 to 1994.  The GX 339-4 images in Fig.~\ref{fig:eb1} and Fig.~\ref{fig:eb2} are the averages of the four individual images during the individual outbursts.  The band 2 image shows clear evidence of emission at the position of GX 339-4.  The black hole transient GS 1354-654 underwent an outburst in 1997-98.  Though it was not as bright as the other black hole transients, the image of GS1354-645 shows a significant peak at the known position of the source (as marked by the `+' sign).

Sco X-1 and Vela X-1 are low-mass x-ray binaries (LMXBs) that are persistent but highly variable sources with soft gamma-ray spectra.  They are moderately bright in band 1, but neither of these sources are seen definitively at energies above 100 keV \cite{ling10}.  GRS 1915+105 and 4U 1700-37 are both LMXBs (GRS 1915+105 containing a black hole and 4U 1700-37 containing a neutron star).  Both sources are moderately bright in band 1 but much fainter in band 2, and neither source is seen above 230 keV \cite{ling10}.  GRS 1009-45 is a transient black hole source that underwent a single outburst in 1993.  At its peak, it was nearly as bright as the Crab in band 1, but much fainter in band 2, and not visible above 230 keV \cite{ling10}.  The extragalactic source NGC 4151 is a relatively faint source that shows up consistently in band 1 images summed over fewer precession cycles, with the central feature of the image consistent with the known position.

The shapes of the features in the images are consistent with expected PSFs based on the source declination as discussed in Section \ref{sec:psf}.  The low declination sources (GRS 1915+105 and Sco X-1) are very elongated in the north-south direction, while the moderate declination sources (Crab, GRO J0422+32, GRO J1719-24, Cyg X-1, and 4U 1700-37) are only sightly elongated with the characteristic `x' shape further away from the peak.  The high declination sources (GX 339-4, GS 1354-645, Vela X-1, and GRS 1009-45) are nearly circular or even elongated in the east-west direction.  Weaker sources such as NGC 4151 and GS 1354-645 have shapes that are more heavily influenced by source confusion (contributions from a limb that has another bright source occulting at the same time), other transient events in the data, e.g. gamma-ray bursts, solar flares, etc., or gaps in the data.

\section{Conclusions and Future Work}
We have briefly described the concept of Earth occultation and the application of the Linear Radon Transform technique to the BATSE LAD data for imaging the gamma-ray sky.  We have shown preliminary images of cosmic gamma-ray sources at energies up to $\sim600$ keV to demonstrate the feasibility and capability of the technique for generating images of the sky in the elusive soft gamma-ray regime, localizing individual sources to $< 1^{\circ}$ accuracy. The images are consistent with the simulated PSFs showing the dependence of the shape of the features on the declination.  

Ongoing work includes quantifying the significance of the features in the images and defining the sensitivity limits for this imaging method.  We are also developing image deconvolution techniques in order to improve the quality of the images and more accurately locate the positions of individual sources, both those that are known and those that are newly discovered.  Our long term goal is to create a map of the entire sky in the four broad energy bands 23-98 keV, 98-230 keV, 230-595 keV, and 595-1800 keV.  This would be the first all-sky map in the $\sim200-2000$ keV region since {\it HEAO} A4. The purpose of the sky map is to look for new sources that were not previously detected, which would then be added to our EBOP input catalog.  The EBOP analysis system will then be rerun to both obtain fluxes for the new sources and to get more accurate fluxes for the previously known sources by reducing the systematic errors currently present due to the unknown sources in the sky.  

This work is directly applicable to the Gamma-Ray Burst Monitor (GBM) on {\it Fermi}, an instrument similar to BATSE.  GBM can also be used to perform Earth occultation measurements, and a program is already underway to monitor gamma-ray sources \cite{hodge09}.  The imaging technique presented here using the LRT can be adapted to use with GBM, allowing new sources to be discovered.  These new sources would be added to the input source catalog for continued monitoring, and the community would be alerted so that follow-up observations could be made at other wavelengths. 

The current method is based on the Linear Radon Transform which requires the image tiles to be relatively small ($\lesssim10^{\circ}$) in order to keep the ``lines'' of integration (i.e. the Earth limbs) approximately linear. The geometry of the Earth occultation technique may be better described as an integration over the sphere with a spherical cap blocked by the Earth. The resulting integral is the spherical cap Radon transform which has many interesting properties, including the ability to create much larger image tiles. By utilizing this new transform in the imaging process, additional information may be extracted from the gamma-ray data to provide sharper images with more accurate flux estimates. Ultimately, we expect this enhanced technique to help identify new faint sources in the BATSE database and to be useful for current missions such as {\it Fermi} as well as future gamma-ray astronomy missions. 

The gamma-ray sky in the 200 keV to 10 MeV region has yet to be fully explored to date.  This is due in part to the complexity and cost of building an appropriate instrument, with wide field of view and good angular resolution for localizing sources to a fraction of a degree.  The combination of a simple instrument, such as BATSE and GBM, with large omni-directional detectors and effective optics provided by the Earth limbs (or moon limbs, if flown in a lunar orbit), coupled to the application of tomographic imaging techniques, such as the LRT as shown in this work, would be an attractive, simple and cost effective option for future gamma-ray missions in the 200 keV to 10 MeV energy region.

\begin{acknowledgments}
The work is funded in part by the NASA SMD AISR Program for project MTool, the NASA USRP Program, and the Caltech SURF Program with support from the JPL Educations Office. The authors wish to thank John Roland for his assistance in generating the images.
\end{acknowledgments}

\end{document}